\documentclass[aps,preprintnumbers,floatfix,amsmath,amssymb,11pt]{revtex4}

\usepackage{graphicx}
\usepackage{dcolumn}
\usepackage{bm}
\usepackage{verbatim}

\begin{document}

\title{Relics of Supersymmetry in Ordinary 1-flavor QCD:\\ Hairpin Diagrams and Scalar-Pseudoscalar Degeneracy}
\author{ Patrick Keith-Hynes and H.B.~Thacker$^{}$ }
\affiliation{
 $^{}$Department of Physics, University of Virginia, Charlottesville, VA 22901 \\ }
\date{\today}


\begin{abstract}
The large-$N_c$ orientifold planar equivalence between $\mathcal{N}=1$ SUSY Yang-Mills theory and ordinary
1-flavor QCD suggests that low-energy quark-gluon dynamics in QCD should be constrained by the supersymmetry of the 
parent theory. One SUSY relic expected from orientifold equivalence is the approximate degeneracy
of the scalar and pseudoscalar mesons in 1-flavor QCD. Here we study the role of the $q\bar{q}$ annihilation
(hairpin) contributions to the meson correlators. These annihilation terms induce mass shifts of
opposite sign in the scalar and pseudoscalar channels, making degeneracy plausible. Calculations of valence
and hairpin correlators in quenched lattice QCD are consistent with approximate degeneracy,
although the errors on the scalar hairpin are large. We also study the role of $q\bar{q}$ annihilation 
in the 1- and 2-flavor Nambu-Jona Lasinio model, where annihilation terms arise
from the chiral field determinant representing the axial $U(1)$ anomaly. Scalar-pseudoscalar degeneracy
for the 1-flavor case reduces
to a constraint on the relative size of the anomalous and non-anomalous 4-fermion couplings.

\end{abstract}
\maketitle


\section{Introduction}

The emerging connections between QCD and string theory provide a promising new theoretical
framework for the study of low energy hadron physics. A particularly interesting outgrowth of these
developments is the large-$N_c$ orientifold
planar equivalence between ${\cal N}=1$ SUSY Yang-Mills theory and ordinary nonsupersymmetric
one-flavor QCD \cite{ASV,ASV_review}. This
correspondence can be described strictly as a field theory to field theory equivalence
but its roots lie in string/gauge duality.  To the extent that orientifold equivalence is reliable at $N_c=3$, it
provides a powerful new source of insight into the chiral dynamics of quarks.  What is particularly
striking about this connection is that the quark of the daughter
theory (one-flavor QCD) is the orientifold projection of a gluino in the parent
theory (${\cal N}=1$ SUSY YM). If this equivalence is even approximately valid
at $N_c=3$, it would expose a deep and surprising role of supersymmetry
in low energy quark-gluon dynamics. The prediction of orientifold equivalence for the
size of the quark condensate in 1-flavor QCD has been compared with lattice results
in Ref. \cite{deGrand06}.
In this paper we consider another of the predictions of orientifold equivalence - the degeneracy
of the lowest lying scalar and pseudoscalar mesons of one-flavor QCD (hereafter 
referred to as $\sigma$ and $\eta'$ respectively) \cite{lat06}.  QCD itself provides no apparent reason to expect such
degeneracy, while in the parent SUSY theory degeneracy is a simple result of the fact that 
the $\sigma$ and $\eta'$ belong to the same Wess-Zumino supermultiplet and that
supersymmery is unbroken.  Such a degeneracy, if valid in one-flavor QCD, would be an unmistakable 
relic of the supersymmetry of the parent theory.

The properties of the scalar and pseudoscalar
mesons of the parent theory are conveniently described in terms of the supersymmetric chiral Lagrangian
constructed by Veneziano and Yankielowicz (VY)\cite{VY}. In this formalism, the scalar
and pseudoscalar mesons are associated with the real and imaginary parts of the 
complex scalar Wess-Zumino field representing gluino-antigluino bound states.  
From the viewpoint of QCD alone, the degeneracy of the scalar and pseudoscalar mesons 
seems somewhat mysterious. In that theory, we are accustomed to thinking in a large-$N_c$ framework
in which the pseudoscalar $\eta'$ meson is a would-be Goldstone boson which acquires a mass 
via the axial anomaly. Lattice calculations have verified the role of the anomaly in generating
the $\eta'$ mass in detail by the study of the quark-line-disconnected ``hairpin
diagram'' contribution to the pseudoscalar correlator \cite{BET0,BET1}. (Here and elsewhere, we use
the term ``hairpin'' to refer to a diagram involving quark-antiquark annihilation or creation where
the quark and antiquark involved emerge from the same hadron.)
In contrast to the would-be Goldstone boson interpretation of the $\eta'$, 
the scalar meson $\sigma$, is, at least naively, a typical P-wave quark-antiquark meson. Phenomenology and lattice
calculations suggest that the corresponding flavor nonsinglet mesons are quite heavy, $\stackrel{>}{\sim}1$ GeV.
The flavor singlet $\sigma$ correlator includes hairpin diagrams along with the valence 
diagram, and, as we will
discuss here, this results in a large negative mass shift of the flavor singlet relative to the 
nonsinglet. The scalar hairpin correlator was calculated
in lattice QCD in Ref. \cite{OZI} as part of a study of the spin-parity structure of the OZI rule.
In fact, the quenched calculations of Refs. \cite{OZI, BET1} 
contain all the Monte Carlo results necessary for testing the orientifold prediction. Although
errors on the scalar hairpin mass shift are sizeable, we will show that within these errors,
the combination of the upward shift of the pseudoscalar mass and the downward shift of the scalar mass,
when evaluated in the 1-flavor theory, brings the two mesons into approximate degeneracy.

Througout this discussion, we adopt a large-$N_c$ view of chiral dynamics, as typified by the Witten-Veneziano
relation, which relates the hairpin-induced mass shift of the pseudoscalar meson to the topological
susceptibility of quenched (pure-glue) QCD. In practical terms, we can take the derivation of the Witten-Veneziano 
relation as consisting of the assumption that the $\eta'$ correlator in full QCD is well approximated
by summing repeated quenched hairpin diagrams, with each hairpin vertex treated as a mass insertion.
Although it would clearly be interesting to compare our
results with the results of a full 1-flavor QCD simulation, within the large $N_c$ framework the quenched
hairpin diagrams actually provide a more direct measurement of the mass shift than the corresponding diagrams
in full QCD, which measure a resummed combination of multiple mass insertions. 
For the pseudoscalar correlator, the
interpretation of the hairpin vertex as a pure mass insertion has been confirmed in 
considerable detail by lattice calculations \cite{BET0,BET1}. 
Less detailed information is known for the scalar hairpin correlator, but we will assume that resummation of 
quenched hairpin vertices also provides a good description of the correlator in the scalar channel,
so that the amputated, quenched hairpin correlator 
is the (mass)$^2$ insertion. It is interesting that, according
to the Monte Carlo results \cite{OZI} the scalar and pseudoscalar hairpin correlators are the only ones which are
large, with the vector and axial-vector hairpins being more than an order of magnitude smaller. This reproduces
the correct phenomenological pattern for the OZI rule, where, for example, the $\omega$-$\rho$ mass splitting is tiny
compared to the $\eta'$-$\pi$ splitting. This entire OZI pattern can in some sense
be seen as a relic of supersymmetry in that it reflects a similar structure in the Veneziano-Yankielowicz chiral
Lagrangian for the parent SUSY theory. In the VY Lagrangian, the auxiliary complex field of the Wess-Zumino multiplet
corresponds to $F^2+iF\tilde{F}$ in the gauge theory. In the low energy chiral Lagrangian, 
this field has no kinetic term, and, when integrated out, 
produces the mass terms for the scalar and pseudoscalar mesons, with the scalar mass arising from integration over
$F^2$ and the pseudoscalar mass arising from integration over $F\tilde{F}$. Thus the gauge fluctuations represented
by the auxiliary fields in the VY Lagrangian result in nonzero hairpin correlators {\it only} in
the scalar and pseudoscalar channels, not in the vector and axial vector channels.
This suggests that the complete pattern of observed OZI-rule violations in hadron phenomenology 
may be plausibly traced to the structure of the chiral Lagrangian in the supersymmetric parent theory. In particular,
the importance of the scalar and pseudoscalar hairpin diagrams, and the relationship between them, reflects
a fundamental SUSY-relic connection between the conformal anomaly and the chiral anomaly in QCD.

Within the large-$N_c$ framework, the quark dynamics leading to scalar-pseudoscalar degeneracy in 1-flavor QCD can be
better understood by generalizing 
to the 2-flavor case.
The separation between valence and hairpin terms is
then obtained by considering appropriate combinations of the isotriplet and isosinglet correlators. In the 2-flavor 
theory we may identify the correlator given by the valence diagrams alone 
with that of the isotriplet meson, which in the pseudoscalar case would be a true Goldstone pion. 
Even in the 1-flavor theory, it is appropriate to
think of the separation between valence and hairpin graphs in the pseudoscalar channel as separating the two types of
chiral symmetry breaking which are taking place: the valence diagrams represent spontaneous $\chi SB$ in the form of
a Goldstone boson, while the hairpin diagrams represent the explicit breaking due to the chiral anomaly. 
In the VY Lagrangian, the pseudoscalar hairpin vertex
is simply the $\eta'$ mass term which arises from the integral over the auxiliary field $F\tilde{F}$.
For the scalar channel, the separation between valence and hairpin contributions to the VY Lagrangian is not 
as obvious. In this case the valence diagrams also contribute to the scalar mass term, and in fact, the mass shift 
from the hairpin graphs turns out to be negative. 

In Section II, we analyze
the Monte Carlo results obtained from scalar and pseudoscalar valence and hairpin correlators and
discuss the significance of these results for the SUSY-relic prediction of scalar-pseudoscalar degeneracy
in 1-flavor QCD. Although the evidence is far from conclusive, the qualitative pattern of valence and
hairpin contributions to the $\sigma$ and $\eta'$ masses which is required for degeneracy is clearly verified
by the Monte Carlo results. We are thus encouraged to investigate the underlying quark interactions which 
lead to this pattern. In Section III we consider this question in a 2-flavor Nambu-Jona Lasinio (NJL) model. This
model plausibly incorporates the chiral dynamics associated with short-range $q\bar{q}$ interactions.
By studying the 2-flavor theory, we deconstruct the meson masses into their valence and hairpin components and
show that the model predicts the same qualitative pattern of masses observed in the Monte Carlo results.
In the 2-flavor NJL model, the axial $U(1)$ anomaly is easily incorporated as an effective 4-fermion vertex
which breaks the $U(2)\times U(2)$ symmetry of the NJL Lagrangian down to $SU(2)\times SU(2)$ \cite{Hatsuda}.
The mass shifts resulting from the axial anomaly can thus be easily determined by a simple application of the
standard Bethe-Salpeter analysis.
We then discuss the effective Lagrangian and mass spectrum of the 2-flavor theory. The effective meson Lagrangian
contains two $[\frac{1}{2},\frac{1}{2}]$
chiral field multiplets (under $SU(2)\times SU(2)$), describing both $I=0$ and $I=1$ mesons. The mass terms
in this theory provide the information needed to show that the non-anomalous $U(2)\times U(2)$ invariant 4-fermion interaction ${\cal L}_s$
represents pure t-channel glue exchange between valence quarks, while the term ${\cal L}_a$ representing
the anomaly is a pure hairpin or $q\bar{q}$ annihilation term.
Within the large-$N_c$ framework, once we have identified the valence-hairpin separation we can generalize the 2-flavor results for the correlators
to any number of flavors by including a factor of $N_f/2$ for each annihilation vertex.

\begin{figure}
\label{fig:plotQCD}
\begin{center}
\includegraphics{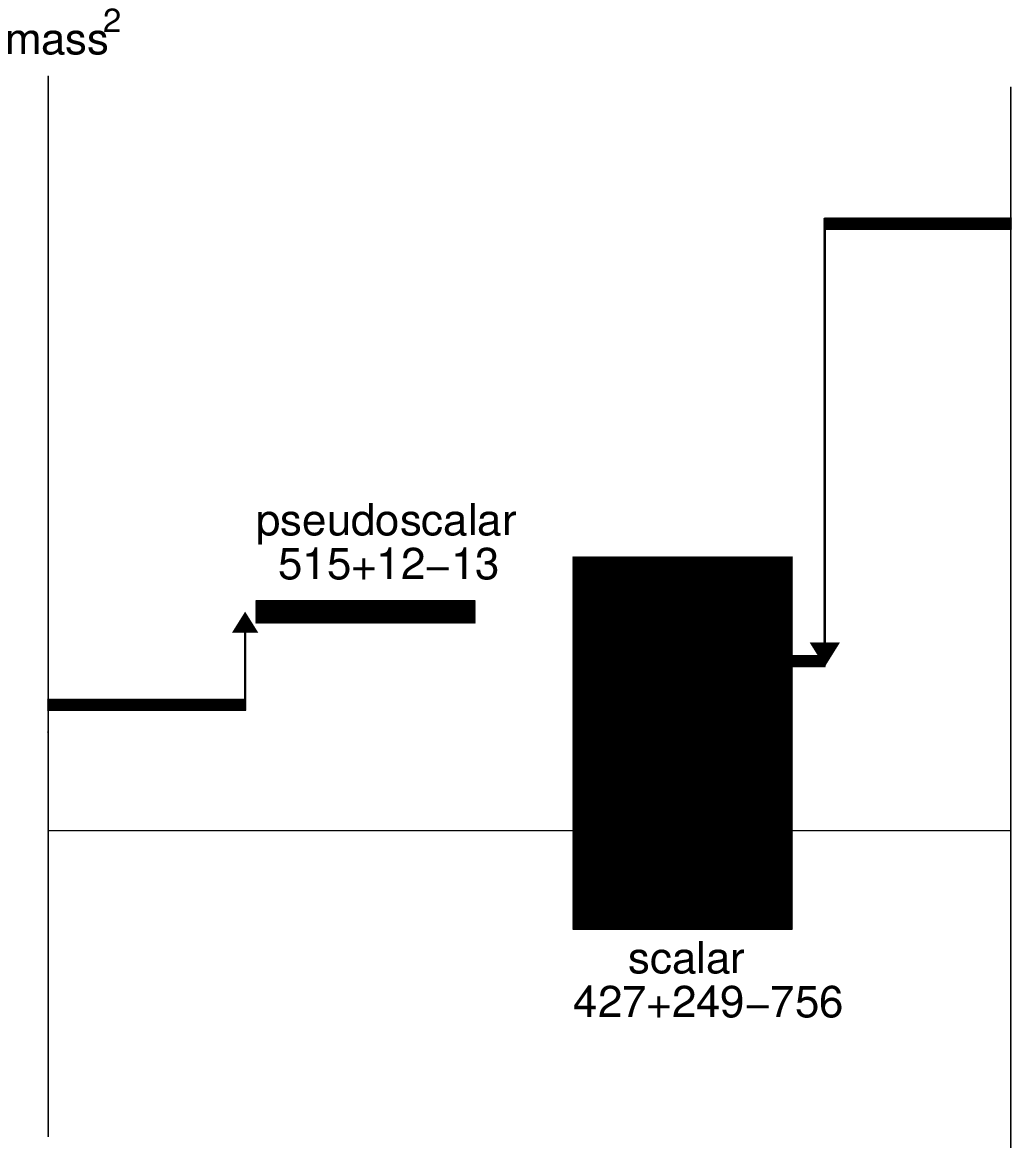}
\caption{Scalar and pseudoscalar mass\textsuperscript{2} for one-flavor QCD (MeV\textsuperscript{2})}
\end{center}
\end{figure}

\begin{table}[h]
\centering
\begin{tabular}{||c|||c|c|c||} \hline \hline
J\textsuperscript{PC}  & Valence Mass\textsuperscript{2} & Mass Shift\textsuperscript{2} & Total Mass\textsuperscript{2}  \\ \hline \hline
$0^{-+}$ & $+[315(6)]^2$ & $+[407(11)]^2$ & $+[515+12-13]^2$ \\ \hline
$0^{++}$ & $+[1416(14)]^2$ & $-[1350(90)]^2$ & $+[427+249-756]^2$ \\ \hline
\end{tabular}
\caption{One-flavor QCD results for scalar and pseudoscalar meson masses. The columns list values in Mev\textsuperscript{2}
for $m_v, m_{hp},$ and $m$, respectively. Errors quoted are statistical only.} 
\label{tab:QCDresults}
\end{table}

\section{Monte Carlo Results}

Ideally one would like to test orientifold equivalence in the chiral limit where it does not
suffer from finite quark mass corrections.
Since the accuracy of the data for the scalar hairpin \cite{OZI} does not
permit a reliable chiral extrapolation, we test the prediction of scalar-pseudoscalar degeneracy 
using one of the lightest quark masses employed in the study of Ref. \cite{OZI} ($m_{\pi}=315(6)$ MeV), 
corresponding to clover improved Wilson fermions
with $C_{sw}=1.57$, hopping parameter $\kappa=.1427$, using 300 quenched gauge configurations generated with
Wilson action at $\beta=5.7$ on a $12^3\times 24$ lattice. 

The valence or nonsinglet mass was extracted from combined exponential fits to smeared and local scalar and pseudoscalar
valence correlators,
\begin{equation}
\int d^3x \langle\bar{u}(x)\Gamma_id(x) \bar{d}(0)\Gamma_i u(0)\rangle \sim \frac{f_i^2}{2m_v} e^{-m_v x_0}
\end{equation}
where $u$ and $d$ are two different flavor Dirac spinors, and $\Gamma_i=1$ or $\gamma_5$. For the scalar valence
correlator a quenched artifact associated with the $\eta'$-$\pi$ intermediate state must be removed from the 
correlator before performing the exponential fit \cite{BET1}. The result for $m_{a_0}$ is in reasonable agreement
with recent full QCD calculations \cite{Prelovsek}. 

The mass shift in the singlet channel is extracted from the
hairpin correlator,
\begin{equation}
\Delta_{hp}(x) = \langle \bar{u}(x)\Gamma_i u(x)\bar{d}(0)\Gamma_id(0)\rangle_c
\end{equation}
Here the subscript $c$ denotes the connected correlator, i.e. for $\Gamma_i=1$ a constant $(VEV)^2$ must 
be subtracted off.
The Monte Carlo results for the quenched hairpin correlators are well fit by a double pole formula
\begin{equation}
\label{eq:doublepole}
\int d^3x \Delta_{hp}(x) \sim \frac{f_i^2 m_{hp}^2}{4m_v^3}(1+m_vx_0)e^{-m_vx_0}
\end{equation}
Upon resummation of multiple hairpin insertions, the mass of the flavor singlet meson is shifted from the 
nonsinglet value $m_v$ to $m$, where
\begin{equation}
m^2 = m_v^2+m_{hp}^2
\end{equation}

   The results of these studies for $\kappa = .1427$ are summarized in Table \ref{tab:QCDresults} and Fig. 1.
The most striking 
finding is the large negative contribution of the hairpin to the mass of the scalar meson, as shown in Fig. 1.  
Although the statistical 
errors in the measurement of the scalar hairpin diagram are large, the size and sign of the scalar mass shift 
make scalar-pseudoscalar mass degeneracy plausible. 

\section{Chiral Lagrangians in ordinary and SUSY QCD} 

The pattern of chiral symmetry breaking in ${\cal N}=1$ SUSY Yang-Mills theory with an $SU(N)$ gauge group
is well understood. The usual description starts
with the chiral anomaly breaking the U(1) chiral symmetry of the gluinos down to $Z_{2N}$. The discrete $Z_{2N}$
chiral symmetry is then spontaneously broken down to $Z_2$ by the appearance of $N$ possible discrete gluino
condensates in the vacuum. In this description, one obtains domain walls between chiral condensates, but 
no Goldstone boson. Nevertheless, one may assume that, for large $N$, the fluctuation of these domain walls 
can be approximately described by a continuously varying order parameter.
Adopting this viewpoint, the Veneziano-Yankielowicz Lagrangian generalizes Witten's 
interpretation of the $\eta'$ meson in QCD as a would-be Goldstone boson \cite{WittenU1Goldstone} 
to the case of ${\cal N}=1$ SUSY YM. 
For the SUSY case, the pseudoscalar meson must then be accompanied in the chiral Lagrangian
by the other members of its supermultiplet, which are degenerate with it since supersymmetry remains unbroken,
even in the presence of a chiral condensate. (The broken chiral $U(1)$ symmetry is
R-symmetry, which is not a supersymmetry transformation, but rather an outer 
automorphism of the SUSY algebra.) In addition to the complex scalar field $\phi$, 
representing the gluino condensate $\bar{\lambda}\lambda$, the basic supermultiplet of the chiral Lagrangian 
includes a fermion field representing gluino-antigluon
bound states and a complex scalar auxiliary field which is identified with $F^2+iF\tilde{F}$ of the gauge
field, where
\begin{equation}
F^2\equiv F^a_{\mu\nu}F^{a\,\mu\nu},\;\;\; F\tilde{F}\equiv\frac{1}{2}\epsilon^{\mu\nu\alpha\beta}F^a_{\mu\nu}F^a_{\alpha\beta}
\end{equation}
The fact that the gauge densities $F^2$ and $F\tilde{F}$ appear in the Lagrangian as auxiliary fields with
no kinetic term amounts to a ``heavy glueball'' approximation in which effects associated with real propagation
of glueballs are neglected. The auxiliary fields in the chiral Lagrangian reflect the relevant vacuum structure 
of the gauge field. 

The construction of the chiral Lagrangian for ${\cal N}=1$ SUSY YM is discussed in detail in the original
reference \cite{VY}. Here we summarize the terms of interest in our discussion. 
The WZ chiral field is given in terms of the gluino condensate order parameter by
\begin{equation}
\label{eq:phi}
\phi = c\bar{\lambda}_R\lambda_L\;, \;\;\;\;\; \phi^*=c\bar{\lambda}_L\lambda_R\;,
\end{equation}
Since we are only discussing the bosonic sector, we will not need the fermionic superpartner
field which is a gluino-gluon composite. We consider only the terms in the Lagrangian that refer to 
the field $\phi$ and the auxiliary field
\begin{equation}
\label{eq:M}
M = -\frac{1}{2}c(F^2+iF\tilde{F})
\end{equation}
In (\ref{eq:phi}) and (\ref{eq:M}), the constant $c$ is given in terms of the renormalization group beta function
\begin{equation}
c = \beta(g)/2g = -\frac{3}{32}g^2N_c/\pi^2 + O(g^4)
\end{equation}
The auxiliary fields are coupled to $\phi$ so that, when they are integrated out, it sets them equal to
\begin{equation}
\label{eq:F2}
cF^2 = -\frac{1}{3}\alpha(\phi^*\phi)^{2/3}\log\left(\phi^*\phi/\mu^6\right)\;+\;\ldots
\end{equation}
and
\begin{equation}
\label{eq:FFdual}
cF\tilde{F} = \frac{1}{3}i\alpha(\phi^*\phi)^{2/3}\log\left(\phi^*/\phi\right)\;+\;\ldots
\end{equation}
where $\;\ldots\;$ represents terms that we are not considering here.
The resulting potential for $\phi$ is
\begin{equation}
\label{eq:V_ss}
{\cal V}_{ss}(\phi) = \frac{1}{9}\alpha(\phi^*\phi)^{2/3}\log(\phi/\mu^3)\log(\phi^*/\mu^3)
\end{equation}
This potential generalizes to the supersymmetric case the large $N$ chiral Lagrangian 
formulation \cite{WittenU1Goldstone}, in which the chiral anomaly is represented by an $\eta'$ mass
term constructed from the log of the chiral field determinant. An exponential parametrization
\begin{equation}
\label{eq:exponential}
\phi = \mu^3 e^{(\sigma+i\eta')/f}
\end{equation}
(with $f=$ pseudoscalar decay constant) shows that the potential (\ref{eq:V_ss}) produces a supersymmetric mass term 
for the scalar and pseudoscalar mesons,
\begin{equation}
{\cal V}_{ss} = \frac{1}{9}\alpha\frac{\mu^4}{f^2}(\sigma^2+\eta'^2)\;+\;\ldots
\equiv \frac{1}{2}m^2_{ss}(\sigma^2+\eta'^2)\;+\;\ldots
\end{equation}

\section{Valence and hairpin masses in the Nambu-Jona Lasinio model}

As discussed in Section II, the lattice determination of the flavor-singlet meson masses requires two separate calculations for the
valence and hairpin terms in the correlator. 
In the pseudoscalar channel, the valence diagram is a massless Goldstone boson propagator, and the hairpin vertex
is a positive mass insertion. By contrast, the scalar 
valence correlator gives a large nonsinglet mass, and the scalar hairpin is a negative mass insertion. 
In this Section, we will investigate the various contributions to the scalar and pseudoscalar masses in
the Nambu-Jona Lasinio (NJL) model. This model has proven to be a useful approximation to QCD which retains much of
the chiral structure of the full theory. There are several modern reviews of the NJL model and its application to
QCD phenomenology\cite{Hatsuda,Bijnens,Klevansky}.
Here we only briefly summarize properties of the model relevant to determining scalar and pseudoscalar masses.
In order to sort out the valence and hairpin contributions in
the chiral Lagrangian of 1-flavor QCD, it is useful to generalize to the 2-flavor theory. 
We define an $I=\frac{1}{2}$ doublet of quark spinors,
\begin{equation}
q = \left(\begin{array}{c} u\\ d \end{array} \right)
\end{equation}
We then write the basic 4-quark interaction of the NJL model in terms of the following
quark bilinears:
\begin{eqnarray}
\label{eq:bilinears}
\Sigma & = & \bar{q}q \\
\Pi_i & = & \bar{q}i\gamma_5 \tau_iq \\
N & = & \bar{q}i\gamma_5 q \\
A_i & = & \bar{q}\tau_iq
\end{eqnarray}
The four-vectors 
\begin{equation}
\label{eq:S}
S = (\Sigma,\vec{\Pi})
\end{equation}
and 
\begin{equation}
\label{eq:P}
P = (N,\vec{A})
\end{equation}
each transform as a $[\frac{1}{2},\frac{1}{2}]$ representation of chiral $SU(2)\times SU(2)$. 
An invariant 4-fermion interaction is given by an arbitrary linear combination of the two 
invariants,
\begin{eqnarray}
\label{eq:invariants}
{\cal L}_1 & = & \Sigma^2 + \vec{\Pi}^2 \\
{\cal L}_2 & = & N^2 + \vec{A}^2
\end{eqnarray}
The $U(2)\times U(2)$ invariant NJL interaction is given by 
\begin{equation}
\label{eq:L_s}
{\cal L}_s = g_s({\cal L}_1 + {\cal L}_2)
\end{equation}
Like the QCD lagrangian, ${\cal L}_s$ is invariant under an axial $U(1)$ transformation 
$q\rightarrow e^{i\alpha\gamma_5/2}q$, or
\begin{equation}
\label{eq:UA1}
(S\pm iP)\rightarrow e^{\pm i\alpha}(S\pm iP)
\end{equation}
The axial $U(1)$ anomaly can be represented by an effective 4-fermion interaction which is invariant
under $SU(2)\times SU(2)$ but violates the $U(1)$ symmetry (\ref{eq:UA1}) \cite{Hatsuda}.
Defining the $2\times 2$ matrix chiral field
\begin{equation}
\Phi_{ab} = \bar{q}_a(1-\gamma_5)q_b
\end{equation}
we take
\begin{equation}
\label{eq:L_anom}
{\cal L}_a = 2g_a \left({\rm Det} \Phi+ {\rm h.c.}\right) = g_a({\cal L}_1 - {\cal L}_2)
\end{equation}
Note that the chiral field can also be written as
\begin{equation}
\label{eq:chfield}
\Phi = \frac{1}{2}\left[(\Sigma+iN)+(\vec{A}+i\vec{\Pi})\cdot\vec{\tau}\right]
\end{equation}

\begin{figure}
\label{fig:self-energy}
\begin{center}
\includegraphics{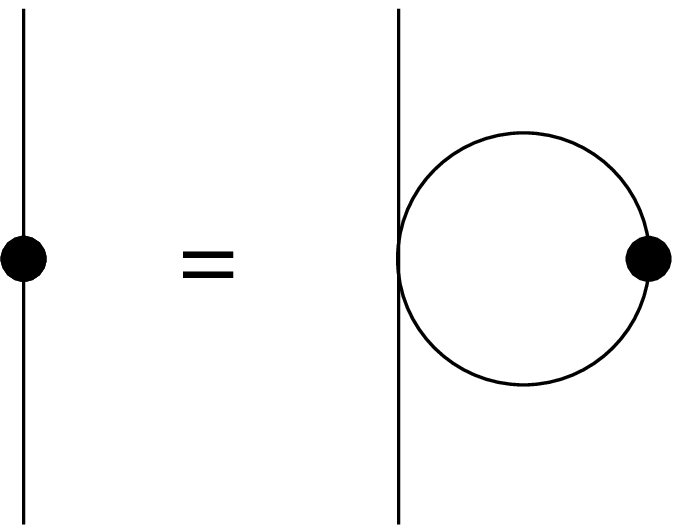}
\caption{Quark self-energy}
\end{center}
\end{figure}

\begin{figure}
\label{fig:bubblesum}
\begin{center}
\includegraphics{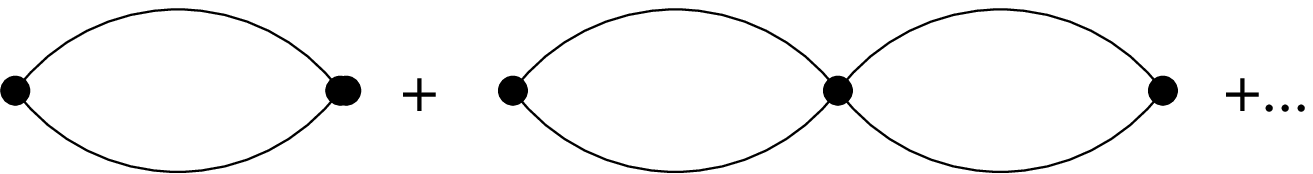}
\caption{Resummed bubble diagrams for the meson correlator}
\end{center}
\end{figure}
Results of the NJL model are given in terms of a constituent quark mass $m_q$, which is the solution to
a self-consistent BCS-type gap equation for the quark self-energy depicted in Fig. 2. 
The mass of the meson in a given spin-parity channel is obtained from the solution to a Bethe-Salpeter equation,
which amounts to a resummed bubble approximation for the meson correlator as shown in Fig 3. A bound state pole corresponds
to a zero of the inverse correlator. Let us first consider the meson masses that are obtained if we neglect
the axial anomaly term and include only the ${\cal L}_{s}$ interaction. Then the gap equation is
\begin{equation}
1 = g_sI_{vac}
\end{equation}
where
\begin{equation}
I_{vac} = \frac{N_c}{2\pi^2}\int_{4m_q^2}^{\Lambda^2} \left(1-\frac{4m_q^2}{s}\right)^{\frac{1}{2}} ds
\end{equation}
The meson correlators are given by a resummation of the one-loop integral in Fig. 3, which reduces
to a dispersion integral of the form
\begin{equation} 
F(q^2,\mu^2) \equiv \frac{N_c}{2\pi^2}\int_{4m_q^2}^{\Lambda^2}\frac{s-\mu^2}{s-q^2}\left(1-\frac{4m_q^2}{s}\right)^{\frac{1}{2}}ds
\end{equation}
Let us denote the scalar meson correlators by $D_{\sigma}$ for the $I=0$ channel and $D_{a_0}$ for the $I=1$ channel,
and the pseudoscalar meson correlators by $D_{\eta'}$ for the $I=0$ channel and $D_{\pi}$ for the $I=1$ channel.
In the theory without the $U_A(1)$ anomaly, the inverse correlators are (dropping an overall constant)
\begin{equation}
D^{-1}_{\eta'}(q^2) = D^{-1}_{\pi}(q^2) = 1-g_sF(q^2,0)
\end{equation}
and
\begin{equation}
D^{-1}_{\sigma}(q^2) = D^{-1}_{a_0}(q^2) = 1-g_sF(q^2,4m_q^2)
\end{equation}
Noting that $F(\mu^2,\mu^2) = I_{vac}$ for any $\mu^2$, the zeroes of the inverse correlators give the meson masses,
\begin{eqnarray}
\label{eq:afmasses}
m^2_{\eta'} & = & m^2_{\pi} = 0 \\
\label{eq:afmasses2}
m^2_{\sigma} & = & m^2_{a_0} = 4m_q^2
\end{eqnarray}

The inclusion of the axial anomaly term ${\cal L}_a$ simply changes the relative strength of the 4-fermion
couplings in the various channels, but the Bethe-Salpeter bubble resummation analysis of ${\cal L}_s+{\cal L}_a$
is similar to the previous discussion. The mass gap equation becomes
\begin{equation}
\label{eq:gap}
1=(g_s+\tilde{g}_a)I_{vac}
\end{equation}
Here $\tilde{g}_a=(1+\frac{1}{2})g_a$ includes a Fierz-transformed cross contraction term.
The inverse correlators for the four channels are now
\begin{eqnarray}
D^{-1}_{\eta'}(q^2) &  = & 1-(g_s-\tilde{g}_a)F(q^2,0) \\
D^{-1}_{\pi}(q^2) & = & 1-(g_s+\tilde{g}_a)F(q^2,0) \\
D^{-1}_{\sigma}(q^2) & = & 1-(g_s+\tilde{g}_a)F(q^2,4m^2) \\
D^{-1}_{a_0}(q^2) & = & 1-(g_s-\tilde{g}_a)F(q^2,4m^2) 
\end{eqnarray}
Note that the coupling constant for the isovector $\pi$ channel is the same as it is for the gap equation (\ref{eq:gap}), so 
again using $F(0,0)=I_{vac}$ we see that the pion remains massless, as it should.
The $\sigma$ channel also has the same coupling as the gap equation, so the mass of the $\sigma$ remains as 
before {\it when expressed in terms of the quark mass},
\begin{equation}
\label{eq:sigmamass}
m_{\sigma}^2 = 4m_q^2
\end{equation}
For the $\eta'$ and $a_0$ masses, the difference in coupling from the gap equation results in an additional mass
shift induced by the $U_A(1)$ anomaly. It is convenient to write the associated Bethe-Salpeter equations 
in terms of the function
\begin{equation}
f(q^2) =\frac{1}{4\pi^2}\int_{4m_q^2}^{\Lambda^2} \frac{1}{s-q^2}\left(1-\frac{4m_q^2}{s}\right)^{\frac{1}{2}} ds
\end{equation}
The resulting equations for the masses of the $\eta'$ and $a_0$ mesons then reduce to
\begin{eqnarray}
\label{eq:etaprimemass}
m_{\eta'}^2f(m_{\eta'}^2) = \frac{1}{N_c}\frac{\tilde{g}_a}{g_s^2-\tilde{g}_a^2}
\end{eqnarray}
and
\begin{equation}
(m_{a_0}^2-4m_q^2)f(m_{a_0}^2) = \frac{1}{N_c}\frac{\tilde{g}_a}{g_s^2-\tilde{g}_a^2}
\end{equation}
These equations can be easily solved numerically. For the present discussion, it is sufficient to consider
the approximation in which the cutoff mass $\Lambda^2$ is large compared to $q^2$ and $m^2$, in which case
\begin{equation}
f(m_{\eta'}^2) \approx f(m_{a_0}^2) \approx f(0) \approx \frac{1}{4\pi^2} \ln\frac{\Lambda^2}{4m_q^2} + {\rm const.}
\end{equation}
In this approximation, the meson masses are
\begin{eqnarray}
\label{eq:fullmasses}
m_{\pi}^2  =  0 ,\;\;&
m_{\eta'}^2  =  m_0^2 \\
m_{\sigma}^2  =  4m_q^2 ,\;\;&
m_{a_0}^2  =  4m_q^2+m_0^2 \label{eq:fullmasses2}
\end{eqnarray}
where we have denoted by $m_0^2$ the mass shift associated with the chiral $U(1)$ anomaly,
\begin{equation}
m_0^2 = \frac{1}{2N_c f(0)}\left(\frac{2\tilde{g}_a}{g_s^2-\tilde{g}_a^2}\right)
\end{equation}
Although it appears that the effect of the $U(1)$ anomaly in the scalar sector is to leave $m_{\sigma}$
unchanged and increase $m_{a_0}$, the interpretation of (\ref{eq:fullmasses2}) must take into account the
fact that the anomalous $g_a$ term also enters the mass gap equation and redefines the quark mass. As shown 
below by a bosonization argument, the actual effect of the $U(1)$ anomaly term in the scalar channel is to leave 
the $a_0$ mass unchanged and lower the $\sigma$ mass. This identifies ${\cal L}_a$ as an annihilation term
which only affects the isosinglet channels.

The pattern of scalar and pseudoscalar masses which emerges from the NJL model can be understood more generally
by considering the effective meson Lagrangian. To proceed, we need to replace the quark bilinears (\ref{eq:bilinears})
by meson fields $\sigma, \vec{\pi}, \eta',$ and $\vec{a}_0$. This can be accomplished by writing an exponential
parametrization of the chiral field $\Phi$, Eq. (\ref{eq:chfield}),
\begin{equation}
\label{eq:chexp}
\Phi = \mu^3\exp[(\sigma+i\eta' + \vec{a_0}\cdot\vec{\tau} +i\vec{\pi}\cdot\vec{\tau})/f]
\end{equation}
If we ignore the $U_A(1)$ anomaly, the $U(2)\times U(2)$ invariant model can be described by an effective
potential of the linear sigma model form,
\begin{equation}
\label{eq:Vsigma}
{\cal V} = -\frac{\alpha}{\mu^2}Tr\Phi^{\dag}\Phi + \frac{\alpha}{2\mu^8}Tr(\Phi^{\dag}\Phi)^2
\end{equation}
Expanding out to quadratic terms in the meson fields, one finds equal mass terms
for the $\sigma$ and $a_0$ fields,
\begin{equation}
\label{eq:chlag}
{\cal V} \approx const.\; + \frac{1}{2}m_1^2(\sigma^2+\vec{a}_0^2)
\end{equation}
where $m_1^2=4\alpha\mu^4/f^2$.
Thus we get the same mass spectrum predicted by the Bethe-Salpeter analysis of the anomaly-free
NJL model, Eqs. (\ref{eq:afmasses})-(\ref{eq:afmasses2}). The degeneracy between isosinglet
and isotriplet mesons in each channel ($m_{\sigma}=m_{a_0}$ and $m_{\eta'}=m_{\pi}=0$) indicates that, in the absence
of the axial anomaly, the $U(2)\times U(2)$ invariant NJL interaction (\ref{eq:L_s}) includes no
hairpin annihilation contribution. The meson propagators obtained in this case are the analog of pure
valence propagators in QCD. The absence of hairpin contributions in ${\cal L}_{s}$ is a specific feature 
of the 4-fermion interaction, and not a consequence of $U(2)\times U(2)$ invariance. To see this, note that in the chiral Lagrangian
(\ref{eq:Vsigma}), we could have included an invariant term of the form
\begin{equation}
(Tr \Phi^{\dag}\Phi)^2
\end{equation}
Although it is $U(2)\times U(2)$ invariant, such a 
term breaks the degeneracy between $\sigma$ and $a_0$ masses, and thus
represents a hairpin annihilation term in the QCD scalar meson propagator. 

To include the axial $U(1)$ anomaly into the effective meson Lagrangian for the 
2-flavor theory, let us first consider the bosonized form of the determinant
interaction (\ref{eq:L_anom}). Using (\ref{eq:chexp}) we find
\begin{equation}
\label{eq:detanomaly}
g_a\left[{\rm Det}\Phi + {\rm h.c}\right] = g_a\mu^6\left[e^{2(\sigma+i\eta')/f}+e^{2(\sigma-i\eta')/f}\right]
\sim 2g_a\mu^6\left[1+2\sigma/f+2(\sigma^2-\eta'^2)/f^2+\ldots\right]
\end{equation}
Thus, the chiral anomaly introduces a negative mass shift for the $\sigma$ along with the expected positive
$\eta'$ mass.
\begin{equation}
\label{eq:scmass}
{\cal V}\rightarrow {\cal V} + \frac{1}{2}m_0^2(\eta'^2-\sigma^2)
\end{equation} 
where $m_0^2=8g_a\mu^6/f^2$.
This reproduces the same pattern of masses (\ref{eq:fullmasses})-(\ref{eq:fullmasses2}) found in both the lattice Monte Carlo
calculations and in the NJL model with the anomaly term included,
\begin{eqnarray}
\label{eq:fullmasses3}
m_{\pi}^2  =  0,\;\; &
m_{\eta'}^2  =  m_0^2 \\
m_{\sigma}^2  =  M^2-m_0^2,\;\; &
m_{a_0}^2  =  M^2
\end{eqnarray}
The bosonic formulation of the
effective Lagrangian makes it more obvious that the effect of the chiral anomaly in the scalar sector is
a negative shift of the $\sigma$ mass, not a positive shift of the $a_0$ mass. In the NJL bound-state analysis, this
requires an interplay between the $a_0$ bound state equation and the gap equation defining the quark mass. 

There is a well-known shortcoming of the form of the chiral 
anomaly represented by the determinant (\ref{eq:L_anom}) \cite{Witten_largeN,DiVecchia,Schechter,Kawarabayashi}. In its bosonized form,
the determinant contains higher order terms in the $\sigma$ and $\eta'$ fields corresponding
to multiply OZI-suppressed meson emission vertices which should be down by additional powers of $1/N_c$. 
The form of the anomaly that satisfies proper OZI counting for large $N_c$ is a pure mass term 
constructed from the log of the determinant,
\begin{equation}
\label{eq:logdetanomaly}
\tilde{\cal L}_a = \frac{1}{2}\tilde{g}_a\mu^6\left[\left(\log{\rm Det}(\Phi/\mu^3)\right)^2\;+\;{\rm h.c.}\right]
= -\frac{1}{2}m_0^2(\eta'^2 - \sigma^2) 
\end{equation}
Comparing this with (\ref{eq:scmass}) we see that the large-$N_c$ form of the chiral 
anomaly, with the appropriate coefficient, produces the same mass shifts as the 4-fermion Det$\Phi$ form
used in the NJL analysis.

\section{Discussion}

From the results of the 2-flavor NJL model, it is straightforward to separate the valence from the hairpin 
contributions to the scalar and pseudoscalar meson correlators. The $U(2)\times U(2)$
invariant interaction ${\cal L}_s$, Eq. (\ref{eq:L_s}), represents a pure valence interaction (i.e. the $q\bar{q}$
interaction arising from t-channel glue exchange) with no $q\bar{q}$ annihilation. The hairpin annihilation vertex
can be identified with the anomaly term ${\cal L}_a$, Eq. (\ref{eq:L_anom}), which breaks $U(2)\times U(2)$ 
down to $SU(2)\times SU(2)$. This separation is inferred from the
fact that, if only ${\cal L}_s$ is included in the Bethe-Salpeter analysis, there is no splitting between 
the $I=0$ and $I=1$ mesons in either channel, identifying this term as a pure valence interaction.
When the ${\cal L}_a$ interaction is included, it effects only the flavor-singlet channels, identifying
this term as a pure $q\bar{q}$ annihilation. 
(Note that in the NJL approximation, even if the actual $q\bar{q}$ interaction is a t-channel exchange
valence graph, a Fierz transformation of the 4-fermion vertex allows 
us to write the spin contractions in the s-channel form, so that the spinor products 
factorize into traces around each loop in the Bethe-Salpeter bubble sum.)

As shown in Section IV, in the meson Lagrangian,
the effect of the anomaly interaction ${\cal L}_a$ in the scalar channel is to introduce a negative
mass shift for the $\sigma$. This agrees at least qualitatively with the Monte Carlo results for the 
hairpin mass shifts. Quantitatively, the chiral Lagrangian analysis predicts equal and opposite (mass)$^2$
shifts, Eq. (\ref{eq:logdetanomaly}), while the Monte Carlo results in Table I and Fig. 1 suggest a much larger shift in
the scalar channel. However, with the large errors in the scalar Monte Carlo data, the significance of this discrepency is
not clear. A more accurate study of the scalar hairpin correlator would clearly be of great interest.

One of the broad goals of the present study is to expose the details of the quark dynamics which underly the
SUSY relic prediction of scalar-pseudoscalar degeneracy. [In this regard, it is interesting that recent work on 
the holographic brane construction of QCD from string theory \cite{Witten98} has derived a generalized NJL model
for quark dynamics starting with a construction of chiral quarks based on $D8$ branes\cite{Harvey}.] 
It is instructive to compare the effective Lagrangian treatment of the NJL model in Section IV with
the supersymmetric Veneziano-Yankielowicz Lagrangian which is presumed by orientifold equivalence to 
approximately describe 1-flavor QCD. 
In the VY Lagrangian, $\sigma$-$\eta'$ degeneracy is a consequence of the fact that the conformal anomaly $\propto F^2$ and the chiral anomaly $\propto F\tilde{F}$
are in the same supermultiplet and hence have equal couplings to the $\log\phi$ terms (c.f. (\ref{eq:F2}) and 
(\ref{eq:FFdual})). Thus in QCD, the statement of scalar-pseudoscalar degeneracy may be interpreted as
a relation between the relative strength of the valence and hairpin interactions.  
Applying the analysis of Section IV to the 1-flavor theory gives an equation for $m_{\eta'}$.
The scalar mass is given by the usual $m_{\sigma}^2=4m^2$. So if we impose the SUSY constraint
\begin{equation}
m_{\sigma}^2=m_{\eta'}^2
\end{equation}
we find that $\sigma$-$\eta'$ degeneracy implies that the $g_s$ and $g_a$ couplings are related by
\begin{equation}
\label{eq:degenerate} 
g_s\int_1^{\lambda}\frac{(2x-1)dx}{\sqrt{x(x-1)}} = \hat{g}_a\int_1^{\lambda}\frac{dx}{\sqrt{x(x-1)}}
\end{equation}
Where $\hat{g}_a=\tilde{g}_a/2=3g_a/4$ and $\lambda=\Lambda^2/4m^2$. This constraint determines the 
strength $g_a$ of the anomaly term needed to give $\sigma$-$\eta'$ degeneracy. We note that, for any 
cutoff $\Lambda^2>m_0^2$, Eq. (\ref{eq:degenerate}) gives $\hat{g}_a<g_s$. For a cutoff close to threshold,
$\Lambda\approx m_0$, it gives $\hat{g}_a\approx g_s$, while for large cutoff $\Lambda >> m_0$, we find
$\hat{g}_a << g_s$. 

The valence-hairpin separation realized in the 2-flavor theory also applies to the 1-flavor theory, and we can identify the
U(1) invariant and U(1) violating 4-fermion interactions as valence and hairpin terms, respectively. Moreover, we can relate this 
separation to a separation between integrals over the $F^2$ and $F\tilde{F}$ auxiliary fields in the VY
Lagrangian. From (\ref{eq:F2}), (\ref{eq:FFdual}), and (\ref{eq:exponential}), it is seen that the sigma
mass comes entirely from the $F^2$ term and the $\eta'$ mass entirely from the $F\tilde{F}$ term.
Thus the $F\tilde{F}$ integration contributes only to the pseudoscalar hairpin 
interaction, while the $F^2$ integration contributes to both the valence and hairpin interactions in
the scalar channel.
We can identify the correct valence-hairpin separation in
the 1-flavor VY effective potential (\ref{eq:V_ss}),
\begin{equation}
\mathcal{V}_{ss} = \mathcal{V}_{val}+\mathcal{V}_{hp}
\end{equation}
The hairpin term can be identified with the large-$N_c$
form of the chiral anomaly (\ref{eq:logdetanomaly}), which for the 1-flavor case is given
in terms of the $U(1)$ chiral field (\ref{eq:exponential}) by 
\begin{equation}
\mathcal{V}_{hp}(\phi) = \frac{1}{18}\alpha(\phi^*\phi)^{2/3}\left[i\log(\phi/\mu^3)\right]^2 \;+\;{\rm h.c.}
\end{equation}
Subtracting this from the supersymmetric potential, we see that the valence term in the VY Lagrangian
is given by
\begin{equation}
\mathcal{V}_{val}(\phi) = \frac{1}{18}\alpha(\phi^*\phi)^{2/3}\left[\log(\phi^*\phi/\mu^6)\right]^2
\end{equation}
This gives the correct separation of mass terms, with ${\cal V}_{val}=m_0^2\sigma^2$ and
${\cal V}_{hp}=\frac{1}{2}m_0^2(\eta'^2-\sigma^2)$.

This work was supported in part by the Department of Energy under grant DE-FG02-97ER41027.


\begin{thebibliography}{99}
\bibitem{Percol} It is tempting to refer to such structure also as a ``percolating
   structure''. However, at this stage it is not clear whether similarities to
   percolation go beyond the unbounded nature of the structure.
\end{thebibliography}
\end{document}